\documentclass[12pt]{iopart}
\pdfoutput=1
\usepackage{color}
\usepackage{graphicx}
\usepackage[pdftex]{hyperref}
\hypersetup{colorlinks=true,citecolor=blue,linkcolor=red,urlcolor=blue}
\usepackage[all]{hypcap}
\begin{document}

\title{Graphene nanoribbons on gold: Understanding superlubricity and edge effects}

\author{L Gigli,$^1$ N Manini,$^2$ A Benassi,$^3$ E Tosatti,$^{1,4}$ A Vanossi,$^{5,1}$ R Guerra$^{1,2}$}
\address{$^1$International School for Advanced Studies (SISSA), Via Bonomea 265, 34136 Trieste, Italy}
\address{$^2$Dipartimento di Fisica, Universit\`a degli Studi di Milano, Via Celoria 16, 20133 Milano, Italy}
\address{$^3$Institute for Materials Science, Max Bergmann Center of Biomaterials, and
Dresden Center for Computational Materials Science (DCCMS), TU Dresden, 01062 Dresden, Germany}
\address{$^4$The Abdus Salam International Centre for Theoretical Physics (ICTP), Strada Costiera 11, 34151 Trieste, Italy}
\address{$^5$CNR-IOM Democritos National Simulation Center, Via Bonomea 265, 34136 Trieste, Italy}

\ead{vanossi@sissa.it, guerra@sissa.it}

\begin{abstract}
We address the atomistic nature of the longitudinal static friction
against sliding of graphene nanoribbons (GNRs) deposited on gold, a
system whose structural and mechanical properties have been recently the
subject of intense experimental investigation.
By means of numerical simulations and modeling we show that the GNR
interior is structurally lubric (``superlubric'') so that the static
friction is dominated by the front/tail regions of the GNR, where the
residual uncompensated lateral forces arising from the interaction with
the underneath gold surface opposes the free sliding.
As a result of this edge pinning the static friction does not grow with
the GNR length, but oscillates around a fairly constant mean value.
These friction oscillations are explained in terms of the GNR-Au(111)
lattice mismatch: at certain GNR lengths close to an integer number of the beat (or
moir\'e) length there is good force compensation and superlubric sliding;
whereas close to half odd-integer periods there is significant pinning
of the edge with larger friction.
These results make qualitative contact with recent state-of-the-art
atomic force microscopy experiment, as well as with the sliding of other
different incommensurate systems.
\end{abstract}

\noindent{\it Keywords\/}: graphene, nanoribbon, superlubricity, friction.

\maketitle

\section{Introduction}

When deposited on a clean flat crystal surface, nano-sized objects
usually provide a well-defined mechanical contact.
For this and other reasons, systems of this kind have been the subject of
extensive nanofriction investigation in recent years both in experiments
\cite{Bardotti96,Dienwiebel04,Mougin08, Dietzel08, Paolicelli08,
 Paolicelli09, Schirmeisen09, Dietzel10a,Dietzel10b, Brndiar11,
 Dietzel13,Feng13}
and in simulations
\cite{Luedtke99, Lewis00, Maruyama04, Pisov07, Filippov08, Bonelli09,
 Guerra10, vanWijk13, Guerra16, Sharp16}.
The gold-graphite interface constitutes an especially smooth and lubric
contact, characterized by tenuous lateral forces, which make it an ideal
workhorse system for basic tribological studies.
In particular the controlled movements of gold and antimony nanoclusters
deposited on the graphite surface provided elegant realizations of
structural lubricity \cite{Erdemir07,VanossiRMP13}, leading to interfaces with
sliding friction forces growing sublinearly with the contact area \cite{Dietzel13,Brndiar11}.
Unfortunately the control over gold nanocluster size and structure is,
at best, statistical, and this limitation prevents a precise control of
the interface geometry and orientation.
More recent work focused on the specular, but better controlled, sliding
of graphitic adsorbates, e.g., in the form of graphene nanoribbons
(GNRs), on gold substrates, and in particular on the Au(111)
surface \cite{Ruffieux16, Kawai16}.
State-of-the-art on-surface synthesis techniques allow the construction
of GNRs of controlled shape on the gold surface \cite{Ruffieux16,
 Jacobberger15}, where size and orientation of the nano object can be
monitored with atomic resolution by means of tip-scanning microscopy in
clean conditions of ultra-high vacuum and low temperature.
In addition, the tip of an atomic force microscope (AFM) can easily induce
displacements of the GNRs on the surface, thanks to the very smooth
GNR-gold interaction.
These forced sliding displacements were exploited by Kawai {\it et
al.} \cite{Kawai16} to probe the frictional properties of this
interface.
That work has provided initial evidence for a weak length dependence of
static friction, defined as the minimum external force needed to
start the longitudinal sliding of the GNRs along their long direction.
That data was however affected by large error bars, which prevented the
clear-cut attribution of the frictional properties.
An intervention of theory is therefore called for to clarify the ideal
frictional behavior to be expected for these systems.

\section{System and Methods}

\begin{figure}\label{fig.sketch}
\centering
\includegraphics[width=0.5\columnwidth]{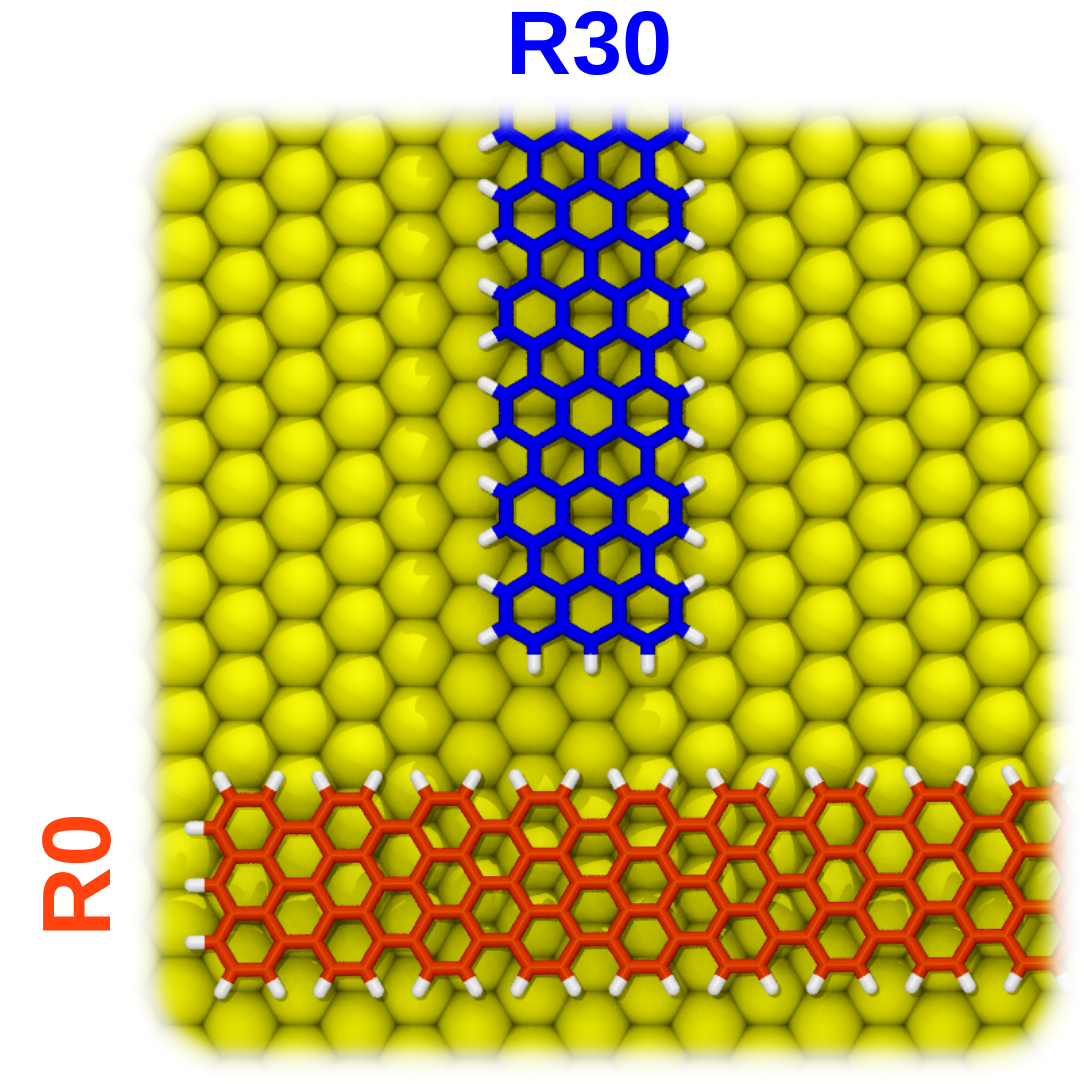}
\caption{
Top-view sketch of the R0 (orange) and R30 (blue) alignments of the GNR
deposited on the Au(111) substrate (yellow).
Broken carbon bonds at front, back, and side edges are passivated
by hydrogen atoms (white).
}
\end{figure}

Mimicking the experimental geometry \cite{Kawai16}, we simulate armchair
GNRs consisting of a stripe of alternating triplets and pairs of carbon
hexagons, as sketched in Figure~\ref{fig.sketch}.
The resulting GNR width is $\simeq 0.7$\,nm, while for the GNR length
$L$ we investigate the experimentally significant range from
$L\simeq 4.2$\,nm (10 unit cells, $N_{\rm C}=140$ C atoms) to
$L\simeq 60.3$\,nm (144 unit cells, $N_{\rm C}=2016$ C atoms).

Our goal is to determine by simulation, and to explain by theory, the static
friction force which resists the longitudinal sliding of these ribbons
on gold.
One complication of real gold (111) surfaces may be represented by
reconstruction, both the primary $(22\times \sqrt{3})$ reconstruction, and the
secondary herringbone long-period one \cite{Woll89, Barth90, Fujita97}.
By ignoring these reconstructions (not always present), we extract the
simplest and fundamental length dependence of friction on
unreconstructed Au(111), represented here by a rigid monolayer
triangular lattice with spacing $a_{\rm Au} = 288.38$\,pm.

We investigate mainly two relative alignments of the GNR with the
substrate: the R0 epitaxial orientation, in which the GNRs are aligned
with the long axis parallel to Au$[1,-2,1]$ ($\theta=0$), and the R30
orientation with the GNRs rotated by $30^\circ$ (or equivalently
$90^\circ$) relative to R0, namely along the Au$[-1,0,1]$ direction
($\theta=30^\circ$), see Figure~\ref{fig.sketch}.
Within our model, depending on the GNR length, the R0 angular alignment
turns out to be only weakly locally stable against global rotations,
with an essentially flat $T=0$ total energy landscape over an angular
range of approximately $\theta \simeq \pm 5^\circ$.
This flatness is probably due to the interplay and mutual cancellation
of the spontaneous interface angular misalignment prescribed by the
Novaco-McTague theory \cite{Novaco77, McTague79, Grey92, MandelliPRL15,
 MandelliPRB15}, that for a full graphene monolayer would demand an
energy minimum at some small $\theta \neq 0$, and the finite size and
elongated shape of the GNR overlayer, favoring perfect alignment at
$\theta = 0$.
In practice, while in the experiment of \cite{Kawai16} the
commonest GNR orientation is the R30 alignment, the R0 arrangement seems
to be favored in the case of graphene flakes grown, with a different
technique, on Au(111) \cite{Wofford12}.
It is also possible that for the GNRs of \cite{Kawai16} the
edges might play a role and tilt the tight energy balance in favor of
the R30 orientation.
Alternatively, the prevalence of R30 might be a purely kinetic effect of
the synthesis method adopted for the GNR \cite{Ruffieux16, Jacobberger15}.
With our model forces, and for large lengths $L$, the R30 orientation is
a local energy minimum of marginally higher energy, approximately
$0.008$\,meV/atom, relative to the R0 orientation.
At any rate, we simulate the static friction of both orientations and
compare the outcomes.

In our model, the GNR atomistic dynamics is determined by the
force field based on the Reactive Empirical Bond Order (REBO) C-C
and C-H interaction potential \cite{Brenner02} as implemented in
LAMMPS \cite{lammps}, plus a 2-body Lennard-Jones (LJ) potential
describing the C-Au interaction.
Following Refs.~\cite{Nie12,Kawai16} we take $\sigma_C=274$\,pm
and $\varepsilon_C=2.5$\,meV for the C-Au LJ parameters, which is best
adapted to model the mutual corrugation energy of graphitic materials
with Au(111) when the latter is represented by a single rigid layer.
As is generally the case in experiment, where broken bonds are
immediately saturated, we assume all broken bonds of the peripheral
C atoms to be H-passivated.
Our tests proved that, even within an empirical force modeling where
electrons are absent, the lack of H passivation, assumed earlier \cite{Kawai16},
would produce significantly shorter C-C bond lengths of the graphitic
edge, thus compromising a realistic moir\'e superstructure
as well as static friction which, as we shall see, is edge-related.
The less important H-Au interaction is also described by a LJ potential
with $\sigma_H=\sigma_C$ and $\varepsilon_H=1.0$\,meV.
We verified that a variation, by a factor of 2 or 3, in $\varepsilon_C$
and $\varepsilon_H$ does not affect the overall trend of friction versus
size, simply rescaling the overall static-friction curves almost rigidly
by the same factor.
We also verified that increasing the number of Au layers representing the
fcc bulk structure of the gold substrate leads to relatively small changes
in the model output, so that these changes can mostly be compensated by a
small adaptation of the LJ C-Au and H-Au parameters to fit the experimentally
observed frequency shifts profiles recorded during the AFM
scans \cite{Kawai16}.

To obtain the fully relaxed initial configuration of each GNR on gold
we run a Langevin simulated thermal annealing by decreasing the target
temperature from $T=50$\,K down to 0\,K in $300$\,ps, at a rate of
$-0.17$\,K/ps, and a subsequent damped relaxation at $T=0$ for a further
$200$\,ps.
To prevent thermally-induced rotations of R0 GNRs during this annealing
protocol, we cancel their out-of-plane angular momentum during the
relaxation dynamics.
In addition, the robustness of the relaxation procedure is tested
against in-plane displacements of the initial center-mass position,
adopting the resulting lowest-energy GNR arrangement as the fully
relaxed initial configuration for the subsequent friction simulations.

Starting from these relaxed configurations, we then evaluate the static
friction by means of zero-temperature molecular-dynamics (MD)
simulations, by applying an adiabatically increasing external force
$F^{\rm tot}$ directed along the GNR main axis.
The force can be applied to the GNR center of mass or, it could be
applied (as a pulling force), to the GNR leading edge.
By directly testing that an edge-driven simulation protocol does not
lead to fundamentally different outcomes, as expected for this kind of
large material stiffness and small length size \cite{Ma15}, we choose
here a uniform center-of-mass driving of the GNR, with the same force
$F=F^{\rm tot}/N$ equally acting on all $N=(N_{\rm C}+N_{\rm H})$
atoms of the adsorbate.
Our protocol is to increase adiabatically $F$ in small incremental
steps $dF \simeq 8\cdot10^{-5}$\,pN/atom,
letting the system structure relax and the kinetic energy be absorbed
after each step before the next.
This relaxation procedure is implemented via a viscous damping rate
$\eta$, whose specific value (here, 2\,ps$^{-1}$), we checked, does not
affect the outcome of the simulated tribological response.
We finally estimate the static friction force $F_s^{\rm tot}$ as the
smallest force $F^{\rm tot}$ leading to a freely sliding configuration,
with the GNR center-mass speed experiencing a frank increase in the pulling
direction exceeding $10^{-4}$\,m/s in simulation.

\section{Static friction}\label{staticfriction:sec}

\begin{figure}\label{fig.statfric}
\centering
\includegraphics[width=0.6\columnwidth]{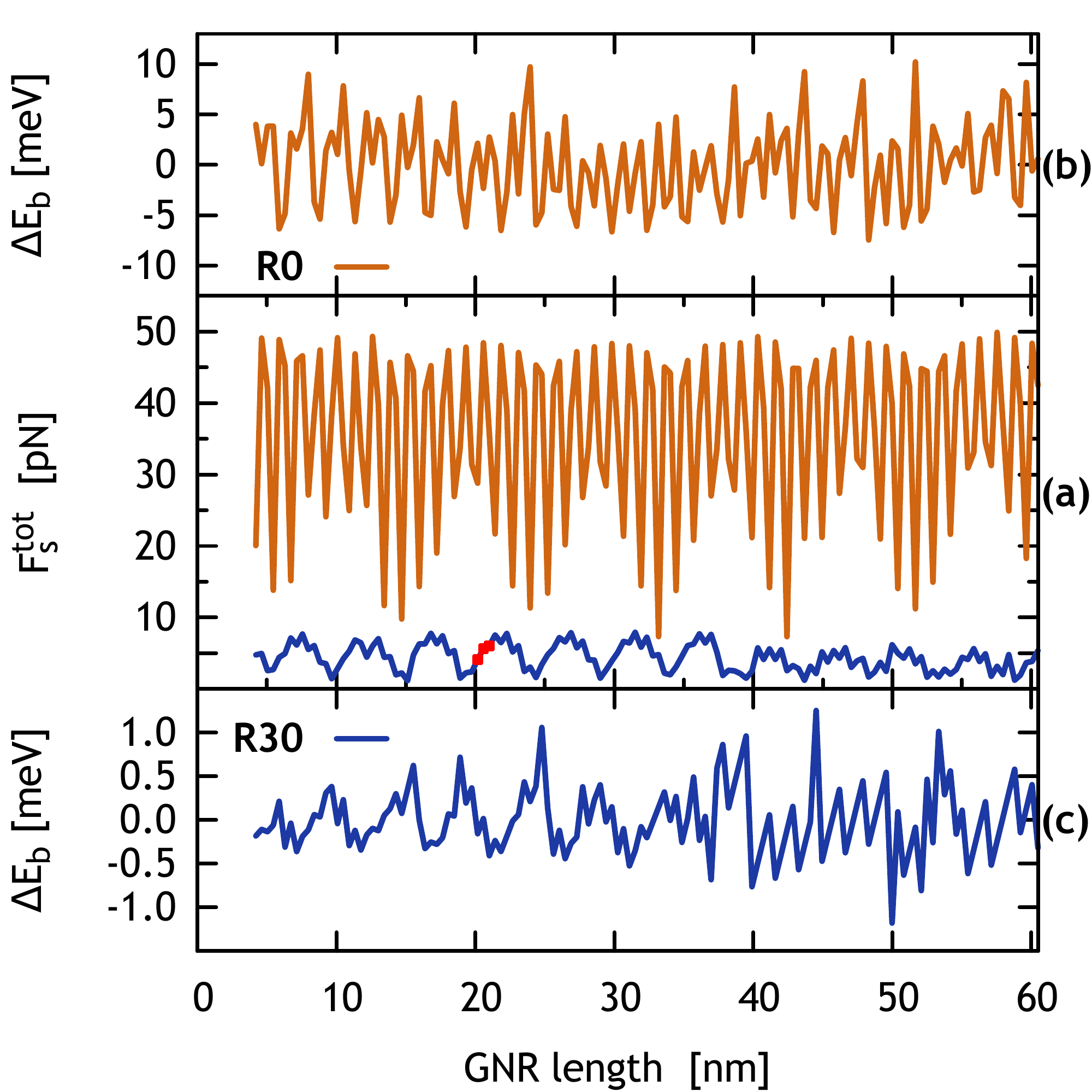}
\caption{
  Dependence on the GNR length $L$ of the computed total static
  friction force (a), and of the variation of the total adsorption
  free energy with respect to its linear fit (b) and (c).
  The GNR orientations are R0 (orange, $\theta=0$) and R30 (blue,
  $\theta = 30^\circ$).
  The linear fits are $E_b=p L + q$, with $p=-345.735$ /
  $-345.891$\,meV/nm and $q=-34.5197$ / $-21.1741$\,meV for R0 / R30
  respectively.
  Red symbols mark the values for the GNRs of Figure~\ref{fig.displacement}.
}
\end{figure}

Figure~\ref{fig.statfric}a reports the total static friction force,
$F_s^{\rm tot}$, as a function of the GNR length for the R0 and R30
orientations.
These results show that the R0-oriented GNRs exhibit a small static
friction ($\sim 20-50$\,pN), which nonetheless is always systematically
larger compared to R30-oriented GNRs ($\sim 5$\,pN).
Interestingly, for both orientations, Figure~\ref{fig.statfric}a displays
periodic-like oscillations of $F_s^{\rm tot}$ as a function of $L$,
without any systematic increase.
Compatible with the experimental observations of \cite{Kawai16}, this
result is consistent with a {\em per atom} $F_s = {\rm const}/L$.
Static friction per particle decreases as a function of the GNR length,
indicating an asymptotically vanishing friction per unit area of contact,
the characteristic hallmark of a superlubric tribological contact.
For the graphene-gold interface, characterized by a fairly weak
graphene-Au(111) interaction, very stiff in-plane C-C bonds, and
mutually mismatched periodicities, one does indeed expect a structurally
lubric behavior, which here emerges even for GNRs, finite-width
graphene ribbons.
The independence of the total static friction for the whole GNR from the
length $L$ shows that there is no term linearly proportional to the contact
area, the very definition of superlubricity \cite{VanossiRMP13}.

\begin{figure*}\label{fig.moire}
\centering
\includegraphics[width=\textwidth]{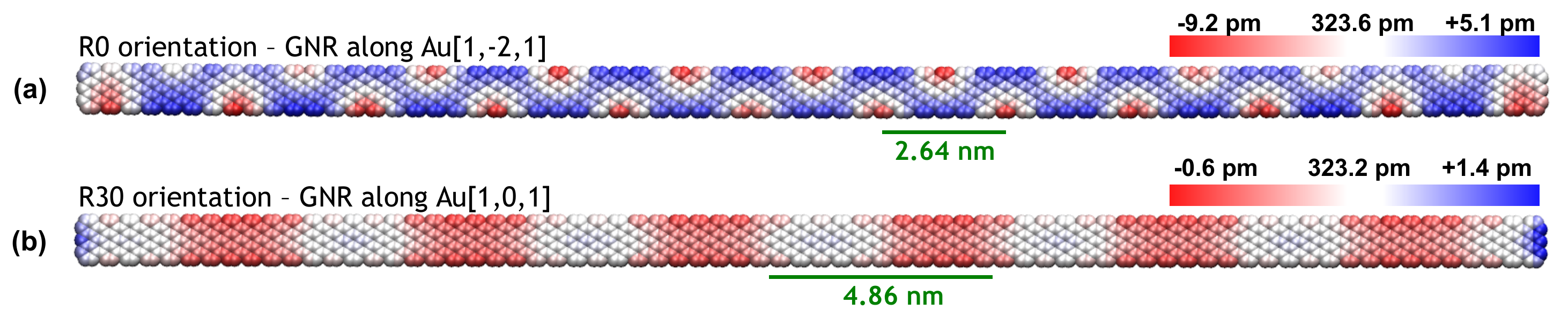}
\caption{
  The relaxed GNR configuration of the 30\,nm-long GNR in the R0 (a)
  and R30 (b) orientations relative to the underlying Au(111) surface
  (not shown).
  The color of the C atoms maps their vertical displacement relative to the
  GNR average height (passivating H atoms are not shown).
  The wavelength $\lambda_m$ of the moir\'e pattern, as extracted from
  the lattice-mismatch value, is indicated for each orientation.
}
\end{figure*}

The observed oscillation of $F_s^{\rm tot}$ can be explained as a
consequence of ribbon-specific moir\'e pattern induced by the
lattice-spacing mismatch.
Figure~\ref{fig.moire} shows typical moir\'e patterns, those of a
$30$\,nm-long relaxed GNR physisorbed along both R0 and R30
orientations.
Here the moir\'e gives rise to, and is mapped by, vertical displacements
$z$ of the individual C atoms.
They are mapped in contrasting colors, exhibiting quasi-periodic color
patterns whose average periods $\lambda_m^{\rm R0}=2.64$\,nm and
$\lambda_m^{\rm R30}=4.86$\,nm are highlighted.
Lower-$z$ regions (red) contribute more, and higher up segments (blue)
contribute less to the total (negative) binding energy $E_b$ of the
physisorbed GNR.
Due to the overall periodicity of the moir\'e-pattern envelop,
in the GNR ``bulk'' interior these oscillations compensate, so that, on
average, $E_b$ grows linearly with $L$.
There remains a residual oscillating contribution $\Delta E_b$ to $E_b$,
which depends on the length of the short terminating section, namely the
GNR part exceeding an integer number of moir\'e average wavelengths $\lambda_m$.
It is precisely this short residual section of the GNR, whose
contribution to energetics does not grow with $L$, that dominates the
static friction force $F_s^{\rm tot}$.
Indeed, periodically in $L$, the GNR terminations explore configurations
ranging from a locally best matching (a minimum of $\Delta E_b$, with
larger $F_s^{\rm tot}$) or compensated (equally weighting red/blue
areas, generating a local maximum of $\Delta E_b$, and a small
$F_s^{\rm tot}$) arrangement.
The anticorrelation of $F_s^{\rm tot}$ with $\Delta E_b$ atom is
illustrated by comparing the panels of Figure~\ref{fig.statfric}.

The moir\'e patterns arise out of the mismatch between the GNR lattice
spacing, $a_{\rm GNR} = 420.00$\,pm, and the spacing of the
gold surface being $a_{\rm Au} = 499.49$\,pm for R0, and $a_{\rm Au}
= 288.38$\,pm for R30.
The resulting mismatch ratios are $\rho_{\rm R0} = 0.8408 \simeq 1$ and
$\rho_{\rm R30} = 1.4564 \simeq 3/2$, respectively.
The wavelengths of the R0 and R30 moir\'e oscillation
patterns \cite{Grey92,Floria96,Vanossi06,Yankowitz12}
are $\lambda_m^{\rm R0} = a_{\rm GNR}/(1-\rho_{\rm R0}) \simeq 2.64$\,nm,
and $\lambda_m^{\rm R30} = a_{\rm GNR}/(3-2\rho_{\rm R30}) \simeq 4.82$\,nm.
For the R0 orientation, the oscillation period should be approximately
$\lambda_m^{\rm R0} \simeq 6\,a_{\rm GNR}$;
however, the main friction oscillation observed in Figure~\ref{fig.statfric}a
exhibits a period $\frac 12 \lambda_m^{\rm R0} \simeq 3\,a_{\rm GNR}$.
The zig-zag nature of the R0 moir\'e patterns (see Figure~\ref{fig.moire})
explains this discrepancy: only one half of the moir\'e period is
relevant for assigning the unbalanced pinned region at the end.
For the R30 case, the moir\'e wavelength $\lambda_m^{\rm R30}$ accounts for
the relatively slow oscillation in $F_s^{\rm tot}$ and $\Delta E_b$ displayed
in the blue curves of Figure~\ref{fig.statfric}.

\begin{figure}\label{fig.compfunc}
\centering
\includegraphics[width=0.6\columnwidth]{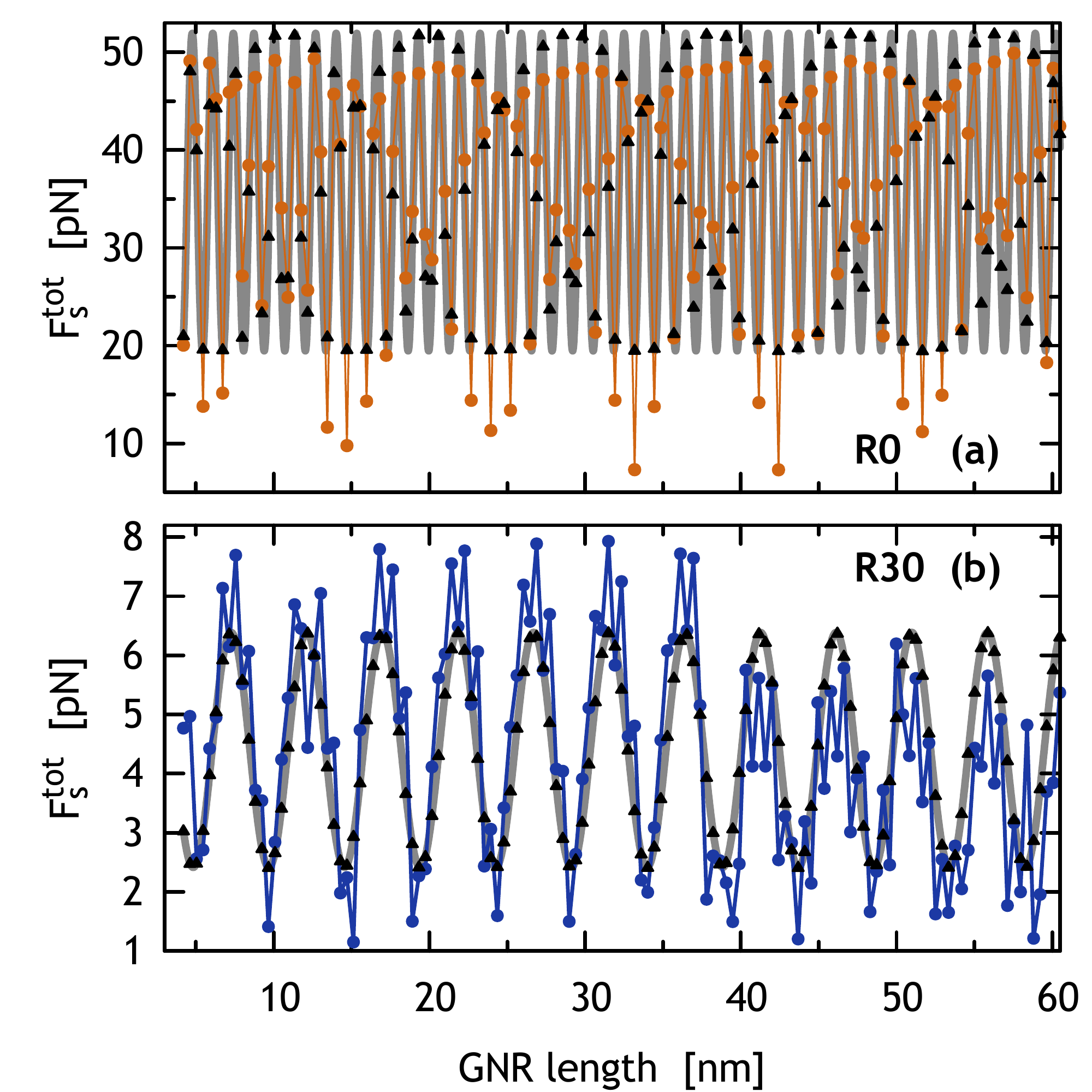}
\caption{
  A comparison of the $L$-dependence of the total static friction force
  $F_s$ (as in Figure~\ref{fig.statfric}a) for (a) R0 and (b) R30
  orientation (circular dots), with the best-fitting curves of
  \eref{eq.compensation} (gray) evaluated at the lengths of actual
  GNRs (triangles).
}
\end{figure}

One can model the dependence of $F_s$ with the GNR length $L$, as if
it was solely related -- as in a purely rigid system -- to the
uncompensated part of the moir\'e pattern of period
$\lambda_m$ \cite{Koren16a,Koren16b}.
In the GNR, whose bulk is superlubric, this part is exclusively related
to the nanoribbon edges -- chiefly the front and the trailing edges, as
the effect of the side edges for longitudinal applied driving turns out
to be negligible.
We can model this uncompensated length, and thus its contribution to
the friction force, with a simple sinusoidal oscillation of period
$\lambda_m$ as a function of $L$ \cite{Koren16b}:
\begin{equation}\label{eq.compensation}
  F(L) = \alpha + \beta
  \sin\!\left( \frac{2\pi L}{\lambda_m} - \delta \right) ,
\end{equation}
where $\alpha$, $\beta$, $\delta$, and $\lambda_m$ are fitting
parameters.
Figure~\ref{fig.compfunc} reports the curves of
\eref{eq.compensation} best-fitting the $F_s$ data of
Figure~\ref{fig.statfric}a.
For the R0 and R30 cases the fit yields $\lambda_m=1.32$\,nm and
$\lambda_m=4.86$\,nm respectively, indeed matching the measured moir\'e
nominal wavelengths (see Figure~\ref{fig.moire}).
The long-wavelength modulations seen for R0 in Figure~\ref{fig.statfric}a
are now explained clearly by the simple sinusoidal fitting function of
Figure~\ref{fig.compfunc}a:
these modulations are the trivial result of a poor sampling rate close
to the Nyquist limit, an aliasing-type effect due to $\frac{1}{2}
\lambda_m^{\rm R0}$ being close, but not quite equal, to $3\,a_{\rm GNR}$.
In the R30 orientation, this effect is not visible since sampling is much
denser, $\lambda_m \gg a_{\rm GNR}$, preventing any aliasing effect.

We note that the close correspondence of the nominal mismatch
with the moir\'e ``periodicities'' in real systems can be perturbed by
relaxation-induced strains and by thermal expansion.
In the case of graphene, which exhibits a large in-plane stiffness, such
strains are severely limited and will mainly influence only the
higher-order approximants, i.e.\ large distances over which deformations
tend to accumulate.
In both cases, R0 and R30, deviations from this simple sinusoidal model
can be attributed to the non-rigid nature of the simulated GNRs, and to
the fact that the uncompensated GNR length does not contribute precisely
as a sine of $L$.
We have repeated the friction simulations with rigid GNRs, with the atomic
reciprocal positions frozen in their configuration fully relaxed in
vacuum, i.e.\ away from the gold surface.
The computed $F_s^{\rm tot}$ values are indeed quite similar, with
typical deviations of the order of 10\%.
One should not take the oscillating function of
\eref{eq.compensation} as a serious model predicting the precise
frictional force experienced by a GNR of a given length, but rather as a
simple interpolating formula expressing the main message of the present
work: the total static friction force needed to set a GNR in motion is
totally edge-related, and does not grow with its length, but rather
oscillates as the residual length of the end section of the GNR
exceeding an integer number of moir\'e wavelengths.

\begin{figure}\label{fig.displacement}
\centering
\includegraphics[width=0.8\columnwidth]{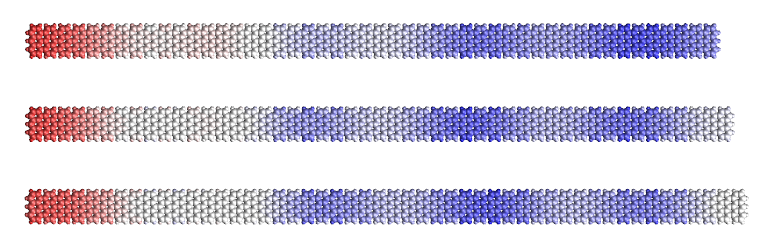}
\caption{
  Colored strain maps for three R30 GNR of $L\simeq20{-}20.8$\,nm
  ($N_{\rm C}=672, 686, 700$), highlighted by red symbols in
  Figure~\ref{fig.statfric}a.
  The map portrays the atomic displacement component (red\,=\,small
  displacement, blue\,=\,large displacement) in the pulling-force
  direction, relative to the fully-relaxed initial configuration evaluated,
  for each GNR, under an uniform external force strength equaling 75\% of
  the corresponding static friction threshold $F_s^{\rm tot}$ as reported
  in Figure~\ref{fig.statfric}a.
  While the central ``bulk'' GNR region clearly exhibits a large mobility,
  reflecting structural lubricity, shear resistance and pinning are
  localized at the nanoribbon ends (here, the GNR tails).
  The calculated static friction threshold turns out to be roughly
  proportional to the extension of this boundary red-blended region of
  the GNR, originating from the corresponding uncompensated part of the
  moir\'e superstructure of the relaxed configurations.
}
\end{figure}

As a confirmation that the main actors responsible for GNR friction are
the end sections, responsible for the uncompensated part of the moir\'e
pattern, Figure~\ref{fig.displacement} shows the atomic displacements of
the GNR atoms in the pulling direction when the applied driving has
reached 75\% of the depinning threshold of each of the three GNRs.
In this picture, the red-colored regions are those resisting shear.
These sections of $\sim \frac 12 \lambda_m \simeq 1{-}2$\,nm near one or
both of the GNR short ends (depending on the moir\'e of starting relaxed
configuration) show up as the main responsible for GNR pinning.
The blue-colored central ``bulk'' GNR region is ready to slide freely if
it was not retained by the stiff elastic interaction with the pinned end
section.
It is therefore established that the short ends, the front end and the
trailing end, represent the source of GNR pinning.
Their relative effectiveness and positioning (front versus tail)
depends on the GNR length in determining how well the matching conditions are
realized at and near these edges.
Similarly to what observed in other investigated tribological systems, yet
with different geometries \cite{Floria96, Braunbook, Cesaratto07, Varini15,
 Pierno15, Koren16a,Koren16b}, we have a final confirmation that the
approximately periodic oscillatory trend of $F_s^{\rm tot}$ as a function
of $L$ shown in Figure~\ref{fig.statfric} is a consequence of the
periodically varying size of the pinning end region.

\section{Depinning}\label{depinning:sec}

As soon as the pulling force exceeds the threshold $F_s^{\rm tot}$, the
GNR starts to move.
While this depinning occurs uneventfully in the R30 orientation, we
observe, interestingly, that the GNR initially aligned at R0 twists away
from the $\theta=0$ orientation, choosing randomly a small (few degrees)
clockwise or counterclockwise angle $\theta$.
While twisting, and subsequently in the depinned state, GNRs start to
slide at an angle $\pm 30^\circ$ away from the pulling force, namely
along one of the Au$[-1,0,1]$ directions, which is the same direction
where R30-aligned GNRs are pulled.
For the shorter GNRs, this twisting at depinning is an essentially rigid
rotation, while it involves small but visible elastic deformations of GNRs
with length $L>30$\,nm.
The reason for this directional locking is that the small twist promotes
the formation of $30^\circ$-oriented flat ``channels'' or ``troughs'' in
the 2D translational energy profile for the center of mass of the GNR,
which is then led to follow these channels where it encounters quite small
barriers against sliding \cite{Tentori17}.
Experimentally, such predicted tendency to directional locking could be
challenging to observe with an AFM, whose force cannot easily be applied
to the center of mass. If observed, it would be remarkable.

\section{Conclusion}

We have presented a study of the static sliding friction of graphene
nanoribbons on a metal surface, performed by classical MD simulations.
Although with no pretense of quantitative predictive power, our model study
appears to capture the essence of their depinning physics.
Paralleling experimental results such as those by Kawai {\it et
 al.} \cite{Kawai16}, we find a small friction, strongly-oscillating and
basically periodic with zero average increase upon increasing GNR length,
supporting superlubricity in this system.
The GNR static friction is entirely edge-related, whereas the GNR
interior is superlubric, much like in a finite-size Frenkel-Kontorova
chain \cite{Floria96, Braunbook, Cesaratto07}, or in other incommensurate
interfaces \cite{Varini15, Pierno15, Koren16a,Koren16b}.
Specifically, simulations allow us to correlate the periodicity of
frictional oscillations to the characteristic length of the moir\'e
pattern, and in particular to the oscillating size of incomplete periods,
namely the front and tail edges, which are responsible for the friction.
This interpretation suggests that GNRs of certain ``magic'' lengths
matching an integer number of moir\'e wavelengths $\lambda_m$ could be
selected for ultra-low-friction applications.

The twist-motion depinning of R0 GNRs induced by a directionally-locked
sliding state and
the dynamical frictional features observed experimentally with the GNRs
either flat on the gold surface or with one end lifted up \cite{Kawai16}
are just a few of the problems calling for future investigation.

{\small
\section*{Acknowledgments}
We acknowledge useful discussions with U.\ Duerig and E.\ Tentori.
Work in Trieste was carried out under ERC Grant 320796 MODPHYSFRICT.
The COST Action MP1303 is also gratefully acknowledged.
}

\vspace{3mm}

\newcommand*{\xdash}[1][3em]{\rule[0.5ex]{#1}{0.55pt}}
\xdash[10em]

\bibliographystyle{iopart-num}
\bibliography{biblio}

\providecommand{\newblock}{}
\begin{thebibliography}{10}
\expandafter\ifx\csname url\endcsname\relax
  \def\url#1{{\tt #1}}\fi
\expandafter\ifx\csname urlprefix\endcsname\relax\def\urlprefix{URL }\fi
\providecommand{\eprint}[2][]{\url{#2}}

\bibitem{Bardotti96}
Bardotti L, Jensen P, Hoareau A, Treilleux M, Cabaud B, Perez A and Aires F~C~S
  1996 {\em Surf. Sci.\/} {\bf 367} 276

\bibitem{Dienwiebel04}
Dienwiebel M, Verhoeven G~S, Pradeep N, Frenken J~W~M, Heimberg J~A and
  Zandbergen H~W 2004 {\em Phys.\ Rev.\ Lett.\/} {\bf 92} 126101

\bibitem{Mougin08}
Mougin K, Gnecco E, Rao A, Cuberes M~T, Jayaraman S, McFarland E~W, Haidara H
  and Meyer E 2008 {\em Langmuir\/} {\bf 24} 1577

\bibitem{Dietzel08}
Dietzel D, Ritter C, M\"onninghoff T, Fuchs H, Schirmeisen A and Schwarz U~D
  2008 {\em Phys. Rev. Lett.\/} {\bf 101} 125505

\bibitem{Paolicelli08}
Paolicelli G, Mougin K, Vanossi A and Valeri S 2008 {\em J. Phys.: Condens.
  Matter\/} {\bf 20} 354011

\bibitem{Paolicelli09}
Paolicelli G, Rovatti M, Vanossi A and Valeri S 2009 {\em Appl. Phys. Lett.\/}
  {\bf 95} 143121

\bibitem{Schirmeisen09}
Schirmeisen A and Schwarz U~D 2009 {\em ChemPhysChem.\/} {\bf 10} 2373

\bibitem{Dietzel10a}
Dietzel D, Feldmann M, Herding C, Schwarz U~D and Schirmeisen A 2010 {\em
  Tribol. Lett.\/} {\bf 39} 273

\bibitem{Dietzel10b}
Dietzel D, M\"onninghoff T, Herding C, Feldmann M, Fuchs H, Stegemann B, Ritter
  C, Schwarz U and Schirmeisen A 2010 {\em Phys. Rev. B\/} {\bf 82} 035401

\bibitem{Brndiar11}
Brndiar J, Turansk\'y R, Dietzel D, Schirmeisen A and \v{S}tich I 2011 {\em
  Nanotechnology\/} {\bf 22} 085704

\bibitem{Dietzel13}
Dietzel D, Feldmann M, Schwarz U, Fuchs H and Schirmeisen A 2013 {\em Phys.
  Rev. Lett.\/} {\bf 111} 235502

\bibitem{Feng13}
Feng X, Kwon S, Park J and Salmeron M 2013 {\em ACS Nano\/} {\bf 7} 1718

\bibitem{Luedtke99}
Luedtke W~D and Landman U 1999 {\em Phys. Rev. Lett.\/} {\bf 82} 3835

\bibitem{Lewis00}
Lewis L~J, Jensen P, Combe N and Barrat J~L 2000 {\em Phys. Rev. B\/} {\bf 61}
  16084

\bibitem{Maruyama04}
Maruyama Y 2004 {\em Phys. Rev. B\/} {\bf 69} 245408

\bibitem{Pisov07}
Pisov S, Tosatti E, Tartaglino U and Vanossi A 2007 {\em J. Phys. Condens.
  Matter\/} {\bf 19} 305015

\bibitem{Filippov08}
Filippov A~E, Dienwiebel M, Frenken J~W~M, Klafter J and Urbakh M 2008 {\em
  Phys. Rev. Lett.\/} {\bf 100} 046102

\bibitem{Bonelli09}
Bonelli F, Manini N, Cadelano E and Colombo L 2009 {\em Eur. Phys. J. B\/} {\bf
  70} 449

\bibitem{Guerra10}
Guerra R, Tartaglino U, Vanossi A and Tosatti E 2010 {\em Nature Mater.\/} {\bf
  9} 634

\bibitem{vanWijk13}
{van Wijk} M, Dienwiebel M, Frenken J and Fasolino A 2013 {\em Phys. Rev. B\/}
  {\bf 88} 235423

\bibitem{Guerra16}
Guerra R, Tosatti E and Vanossi A 2016 {\em Nanoscale\/} {\bf 8} 11108

\bibitem{Sharp16}
Sharp T, Pastewka L and Robbins M 2016 {\em Phys. Rev. B\/} {\bf 93} 121402

\bibitem{Erdemir07}
Erdemir A and Martin J~M (eds) 2007 {\em Superlubricity\/} (Elsevier,
  Amsterdam,)

\bibitem{VanossiRMP13}
Vanossi A, Manini N, M, Urbakh, Zapperi S and Tosatti E 2013 {\em Rev. Mod.
  Phys.\/} {\bf 85} 529

\bibitem{Ruffieux16}
Ruffieux P, Wang S, Yang B, S\'anchez-S\'anchez C, Liu J, Dienel T, Talirz L,
  Shinde P, Pignedoli C~A, Passerone D, Dumslaff T, Feng X, M\"ullen K and
  Fasel R 2016 {\em Nature\/} {\bf 531} 489

\bibitem{Kawai16}
Kawai S, Benassi A, Gnecco E, S\"ode H, Pawlak R, Feng X, M\"ullen K, Passerone
  D, Pignedoli C~A, Ruffieux P, Fasel R and Meyer E 2016 {\em Science\/} {\bf
  351} 957

\bibitem{Jacobberger15}
Jacobberger R~M, Kiraly B, Fortin-Deschenes M, Levesque P~L, McElhinny K~M,
  Brady G~J, Delgado R~R, Roy S~S, Mannix A, Lagally M~G, Evans P~G, Desjardins
  P, Martel R, Hersam M~C, Guisinger N~P and Arnold M~S 2015 {\em Nature
  Comm.\/} {\bf 6} 8006

\bibitem{Woll89}
W\"oll C, Chiang S, Wilson R~J and Lippel P~H 1989 {\em Phys. Rev. B\/} {\bf
  39} 7988

\bibitem{Barth90}
Barth J~V, Brune H, Ertl G and Behm R~J 1990 {\em Phys. Rev. B\/} {\bf 42} 9307

\bibitem{Fujita97}
Fujita D, Amemiya K, Yakabe T, Nejoh H, Sato T and Iwatsuki M 1997 {\em Phys.
  Rev. Lett.\/} {\bf 78} 3904

\bibitem{Novaco77}
Novaco A and McTague J 1977 {\em Phys.\ Rev.\ Lett.\/} {\bf 38} 1286

\bibitem{McTague79}
McTague J and Novaco A 1979 {\em Phys. Rev. B\/} {\bf 19} 5299

\bibitem{Grey92}
Grey F and Bohr J 1992 {\em Europhys. Lett.\/} {\bf 18} 717

\bibitem{MandelliPRL15}
Mandelli D, Vanossi A, Manini N and Tosatti E 2015 {\em Phys. Rev. Lett.\/}
  {\bf 114} 108302

\bibitem{MandelliPRB15}
Mandelli D, Vanossi A, Invernizzi M, Paronuzzi S, Manini N and Tosatti E 2015
  {\em Phys. Rev. B\/} {\bf 92} 134306

\bibitem{Wofford12}
Wofford J~M, Starodub E, Walter A~L, Nie S, Bostwick A, Bartelt N~C, Th\"urmer
  K, Rotenberg E, McCarty K~F and Dubon O~D 2012 {\em New J. Phys.\/} {\bf 14}
  053008

\bibitem{Brenner02}
Brenner D~W, Shenderova O~A, Harrison J~A, Stuart S~J, Ni B and Sinnott S~B
  2002 {\em J. Phys.: Condens. Matter\/} {\bf 14} 783

\bibitem{lammps}
Plimpton S 1995 {\em J. Comput. Phys.\/} {\bf 117} 1

\bibitem{Nie12}
Nie S, N, Bartelt C, Wofford J~M, Dubon O~D, McCarty K~F and Th\"urmer K 2012
  {\em Phys. Rev. B\/} {\bf 85} 205406

\bibitem{Ma15}
Ma M, Benassi A, Vanossi A and Urbakh M 2015 {\em Phys. Rev. Lett.\/} {\bf 114}
  055501

\bibitem{Floria96}
Flor\'\i{}a L and Mazo J 1996 {\em Adv. Phys.\/} {\bf 45} 505

\bibitem{Vanossi06}
Vanossi A, Manini N, Divitini G, Santoro G and Tosatti E 2006 {\em Phys. Rev.
  Lett.\/} {\bf 97} 056101

\bibitem{Yankowitz12}
Yankowitz M, Xue J, Cormode D, Sanchez-Yamagishi J, Watanabe K, Taniguchi T,
  Jarillo-Herrero P, Jacquod P and LeRoy B 2012 {\em Nature Phys.\/} {\bf 8}
  382

\bibitem{Koren16a}
Koren E and Duerig U 2016 {\em Phys. Rev. B\/} {\bf 93} 201404

\bibitem{Koren16b}
Koren E and Duerig U 2016 {\em Phys. Rev. B\/} {\bf 94} 045401

\bibitem{Braunbook}
Braun O~M and Kivshar Y~S 2004 {\em The Frenkel-Kon\-torova Model: Concepts,
  Methods, and Applications\/} (Sprin\-ger, Berlin)

\bibitem{Cesaratto07}
Cesaratto M, Manini N, Vanossi A, Tosatti E and Santoro G~E 2007 {\em Surf.
  Sci.\/} {\bf 601} 3682

\bibitem{Varini15}
Varini N, Vanossi A, Guerra R, Mandelli D, Capozza R and Tosatti E 2015 {\em
  Nanoscale\/} {\bf 7} 2093

\bibitem{Pierno15}
Pierno M, Bruschi L, Mistura G, Paolicelli G, di~Bona A, Valeri S, Guerra R,
  Vanossi A and Tosatti E 2015 {\em Nature Nanotech.\/} {\bf 10} 714

\bibitem{Tentori17}
Tentori E 2017 {\em Energetica di un nastro di grafene nanoscopico depositato
  su Au(111), al variare di posizione e allineamento angolare\/} Master's
  thesis University Milan
  \urlprefix\url{http://materia.fisica.unimi.it/manini/theses/tentori.pdf}

\end{thebibliography}

\end{document}